\documentclass[11pt,a4paper]{article}
\usepackage{jcappub}

\usepackage{amsmath,amssymb}
\usepackage{graphicx}

\newcommand{\bk}{{\boldsymbol k}}

\newcommand{\bx}{{\boldsymbol x}}
\newcommand{\bB}{{\boldsymbol B}}

\newcommand{\beq}{\begin{equation}}
\newcommand{\eeq}{\end{equation}}
\newcommand{\ber}{\begin{eqnarray}}
\newcommand{\eer}{\end{eqnarray}}

\def\ini{{\rm in}}

\def\re{{\rm e}}

\def\rR{{\rm R}}
\def\rL{{\rm L}}

\def\ave#1{\left\langle #1 \right\rangle}

\def\rot{{\rm rot\,}}

\title{Magnetic fields and chiral asymmetry in the early hot universe}

\author[a]{Maksym Sydorenko}
\author[b,c]{Oleksandr Tomalak}
\author[a,c]{and Yuri Shtanov}

\affiliation[a]{Bogolyubov Institute for Theoretical Physics,  03680 Kiev, Ukraine} %
\affiliation[b]{Institut f\"ur Kernphysik, Johannes Gutenberg Universit\"at,  55128 Mainz, Germany} %
\affiliation[c]{Department of Physics, Taras Shevchenko National University,  03022 Kiev, Ukraine} %

\emailAdd{maxsydorenko@gmail.com}
\emailAdd{tomalak@uni-mainz.de}
\emailAdd{shtanov@bitp.kiev.ua}

\abstract{In this paper, we study analytically the process of external generation and subsequent
free evolution of the lepton chiral asymmetry and helical magnetic fields in the early
hot universe. This process is known to be affected by the Abelian anomaly of the
electroweak gauge interactions.  As a consequence, chiral asymmetry in the fermion
distribution generates magnetic fields of non-zero helicity, and vice versa.  We take
into account the presence of thermal bath, which serves as a seed for the development of
instability in magnetic field in the presence of externally generated lepton chiral
asymmetry. The developed helical magnetic field and lepton chiral asymmetry support each
other, considerably prolonging their mutual existence, in the process of `inverse
cascade' transferring magnetic-field power from small to large spatial scales. For
cosmologically interesting initial conditions, the chiral asymmetry and the energy
density of helical magnetic field are shown to evolve by scaling laws, effectively
depending on a single combined variable. In this case, the late-time asymptotics of the
conformal chiral chemical potential reproduces the universal scaling law previously found
in the literature for the system under consideration. This regime is terminated at lower
temperatures because of scattering of electrons with chirality change, which
exponentially washes out chiral asymmetry. We derive an expression for the termination
temperature as a function of the chiral asymmetry and energy density of helical magnetic
field.}

\keywords{primordial magnetic fields, leptogenesis}

\arxivnumber{1607.04845}

\begin{document}
\maketitle \flushbottom


\section{Introduction}

Observations have established the presence of magnetic field of various magnitudes and on
various spatial scales in our universe.  Galaxies such as Milky Wave contain regular
magnetic fields of the order of $\mu$G, while coherent fields of the order of $100~\mu$G
are detected in distant galaxies \cite{Bernet:2008qp, Wolfe:2008nk}. There is a strong
evidence for the presence of magnetic field in intergalactic medium, including voids
\cite{Tavecchio:2010mk, Ando:2010rb, Neronov:1900zz, Dolag:2010}, with strengths
exceeding $\sim 10^{-15}$~G\@.  This supports the idea of cosmological origin of magnetic
fields, which are subsequently amplified in galaxies, probably by the dynamo mechanism
(see reviews \cite{Widrow:2002ud, Kandus:2010nw, Durrer:2013pga, Subramanian:2015lua}).

The origin of cosmological magnetic field is a problem yet to be solved, with several
pos\-sible mechanisms under discussion. These can broadly be classified into inflationary
and post-inflationary scenarios.  Both types still face problems to overcome:
inflationary magnetic fields are constrained to be rather weak, while those produced
after inflation typically have too small coherence lengths (see \cite{Widrow:2002ud,
Kandus:2010nw, Durrer:2013pga, Subramanian:2015lua} for a review of these mechanisms and
assessment of these difficul\-ties). It should also be noted that generation of helical
hypermagnetic field prior to the elec\-troweak phase transition may explain the observed
baryon asymmetry of the universe \cite{Fujita:2016igl, Kamada:2016eeb}.

One of the mechanisms of generation of cosmological magnetic fields which is currently
under scrutiny is based on the Abelian anomaly of the electroweak interactions
\cite{Joyce:1997uy, Frohlich:2000en, Frohlich:2002fg}. If the difference between the
number densities of right-handed and left-handed charged fermions in the early hot
universe happens to be non-zero (as in the leptogenesis scenario involving physics beyond
the standard model; see \cite{Davidson:2008bu, Fong:2013wr} for reviews), then a specific
instability arises with respect to generation of helical (hypercharge) magnetic field.
The generated helical magnetic field, in turn, is capable of supporting the fermion
chiral asymmetry, thus prolonging its own existence to cosmological temperatures as low
as tens of MeV \cite{Boyarsky:2011uy}.  In this process, magnetic-field power is
permanently transferred from small to large spatial scales (the phenomenon known as
`inverse cascade'). Further investigation of the general properties of the regime of
inverse cascade revealed certain universal scaling laws in its late-time asymptotics
\cite{Hirono:2015rla, Xia:2016any, Yamamoto:2016xtu}.

In this paper, we study analytically the process of generation of helical magnetic field
in the early hot universe by an unspecified external source of lepton chiral asymmetry.
Helical magnetic field is produced due to the presence of thermal background, which we
extrapolate to all spatial scales, including the super-horizon scales.\footnote{The
spectral properties of magnetic fields on superhorizon spatial scales depend on a
concrete model of generation of primordial magnetic fields (see \cite{Kandus:2010nw,
Durrer:2013pga, Subramanian:2015lua} for recent reviews).} We consider a simple model of
generation of magnetic field which assumes that the source of chiral anomaly maintains a
constant value of the (conformal) chiral chemical potential of charged leptons. After
generation of magnetic field of near maximal helicity, its evolution is traced in the
absence of the external source of lepton chiral asymmetry. In this case, the helical
magnetic field and the lepton chiral asymmetry are mutually sustained (decaying slowly)
by quantum anomaly until temperatures of the order of tens of MeV, with magnetic-field
power being permanently transferred from small to large spatial scales in the regime of
inverse cascade. We obtain analytic expressions describing the evolution of the lepton
chiral chemical potential and magnetic-field energy density.  The evolution of both these
quantities exhibits certain scaling behavior, effectively depending on a single combined
variable. In this case, the late-time asymptotics of the chiral chemical potential
reproduces the universal scaling law previously found in the literature for the system
under investigation \cite{Hirono:2015rla, Xia:2016any, Yamamoto:2016xtu}. As the
temperature drops down because of the cosmological expansion, the processes of lepton
scattering with the change of chirality (the so-called chirality-flipping processes)
start playing important role, eventually leading to a rapid decay of the lepton chiral
asymmetry.  We give an analytic expression for the temperature at which this happens,
depending on the initially generated values of the magnetic-field energy density and
lepton chiral asymmetry.

\section{Helical magnetic fields}
\label{sec:helical}

A spatially flat expanding universe filled by relativistic matter is conveniently
described in the comoving conformal coordinate system $(\eta, \bx)$ with the conformal
time $\eta$ and scale factor $a (\eta)$ entering the metric line element as $ds^2 = a^2
(\eta) \left( d \eta^2 - d \bx^2 \right)$.  By rescaling the conformal coordinates
$(\eta, \bx)$, one can suitably normalize the scale factor $a (\eta)$.

A divergence-free statistically homogeneous and isotropic cosmological magnetic field has
the following general Fourier representation of the two-point correlation
function\footnote{The quantities $B_i$ are the components of the so-called comoving
magnetic field, which is related to the observable magnetic field strength $\bB_{\rm
obs}$ by the equation $\bB = a^2 \bB_{\rm obs}$. The spatial vector indices are treated
by using the Kronecker delta-symbol, and their position does not matter.}
\cite{Caprini:2003vc}:
\beq \label{correl}
\ave{B_i^{} (\bk) B_j^* (\bk')} = \left( 2 \pi \right)^3 \delta \left(\bk - \bk' \right)
\left[ P_{ij} (\bk) S(k) + i \epsilon_{ijs} \hat k_s A(k) \right] \, ,
\eeq
where $\hat k_i = k_i / k$, $ P_{ij} = \delta_{ij} - \hat{k}_i \hat{k}_j$ is the
symmetric projector to the plane orthogonal to $\bk$, and $ \epsilon_{ijk} $ is the
normalized totally antisymmetric tensor.

It is useful to introduce the helicity components $B_\pm (\bk)$ of the magnetic field via
\beq
B_i (\bk) = B_+ (\bk) \re^+_i (\bk) + B_- (\bk) \re^-_i (\bk) \, ,
\eeq
where the complex basis $ \re^\pm_i = \frac{1}{\sqrt{2}}\left( \re^1_i \pm i \re^2_i
\right)$ is formed from a right-handed (with respect to the orientation $\epsilon_{ijk}$)
and orthonormal (with respect to the metric $\delta_{ij}$) basis ${\bf e}^{1} (\bk)$,
${\bf e}^{2} (\bk)$, ${\bf e}^{3} (\bk) = \bk /k$.  The coefficients of the symmetric and
antisymmetric parts of the correlation function are then expressed through these
components as follows:
\ber \label{S}
\ave{B_-^{} (\bk) B_-^* (\bk') + B_+^{} (\bk) B_+^* (\bk')} &=&
2 \left( 2 \pi \right)^3 \delta \left(\bk - \bk' \right) S (k) \, , \\
\ave{B_-^{} (\bk) B_-^* (\bk') - B_+^{} (\bk) B_+^* (\bk')} &=& 2 \left( 2 \pi \right)^3
\delta \left(\bk - \bk' \right) A (k) \, . \label{A}
\eer
We note an obvious constraint $|A (k)| \leq S (k)$.

The spectrum $A (k)$ of the magnetic-field correlation function characterizes the
difference in the power between the left-handed and right-handed magnetic field, i.e.,
its helicity. The spectrum $S (k)$ characterizes the magnetic field energy density. In
the case of so-called maximally helical magnetic field, one has $ |A(k)| = S(k)$, and
magnetic field is dominated by its left-handed or right-handed part, depending on the
sign of $A (k)$.

In this paper, we consider the effects of Abelian anomaly in the presence of spatially
homogeneous chiral asymmetry.\footnote{Effects of spatial inhomogeneity in chiral
relativistic plasma were under investigation in \cite{Boyarsky:2015faa, Gorbar:2016qfh}.}
In this case, the evolution of the comoving magnetic field in conformal coordinates in
cosmic plasma with high conductivity $\sigma$ takes the form
\beq \label{magev}
\frac{\partial \bB}{\partial \eta} = \frac{1}{\sigma_c} \nabla^2 \bB - \frac{\alpha
\Delta \mu}{\pi \sigma_c} \rot \bB \, ,
\eeq
where $\Delta \mu \equiv a \left(\mu_\rL - \mu_\rR \right)$ is the spatially homogeneous
difference between the (conformal) chemical potentials of the left-handed and
right-handed charged leptons, $\sigma_c \equiv a \sigma \approx {\rm const}$
\cite{Turner:1987bw, Baym:1997gq} characterizes the plasma conductivity, and $\alpha
\approx 1/137$ is the fine structure constant.  The last term in equation (\ref{magev})
is connected with the anomalous current in Maxwell's equations \cite{Joyce:1997uy,
Frohlich:2000en, Frohlich:2002fg, Vilenkin:1980fu, Redlich:1984md, Fukushima:2008xe,
Tashiro:2012mf}.

Using equations (\ref{correl}) and (\ref{magev}), one can obtain the following system of
equations for the spectra $S (k, \eta)$ and $A (k, \eta)$ (see \cite{Boyarsky:2011uy}):
\ber
\frac{\partial S}{\partial \eta}  &=& - \frac{2 k^2}{\sigma_c} (S - S_{\rm eq}) + \frac{2
\alpha k}{\pi \sigma_c} \Delta \mu A \, ,
\label{dotS} \\
\frac{\partial A}{\partial \eta}  &=& - \frac{2 k^2}{\sigma_c} A + \frac{2 \alpha k}{\pi
\sigma_c} \Delta \mu S \, . \label{dotA}
\eer
In equation (\ref{dotS}), we have added a term with the thermal equilibrium distribution
\beq \label{Seq}
S_{\rm eq} (k, \eta) = \frac{k }{e^{k/aT} - 1} \, ,
\eeq
whose role is to ensure relaxation of the spectral energy distribution $S$ to its
equilibrium value $S_{\rm eq}$ rather than to zero. This mechanism will not work in the
long-wavelength domain $k \lesssim 1/\eta$, which is not causally connected in the
expanding hot unverse. This, however, will be of no practical importance, since the
anomalous term in equation (\ref{dotS}) will dominate in this spectral region.  The
initial spectra in the domain of small values of $k$ will also depend on their
cosmological origin.  We do not consider this issue in the present paper, assuming the
initial spectrum to be given by (\ref{Seq}) on all scales.

In an early radiation-dominated universe expanding adiabatically with the bulk matter in
local thermal equilibrium, the entropy density $(a T)^3 g_*$ remains constant. Here,
$g_*$ is the number of relativistic degrees of freedom $g_*$ in thermal equilibrium.  In
the range of temperatures $80~{\rm GeV} < T < 150~{\rm MeV}$, the value of $g_*$ changes
insignificantly from about 86 to 72, and at the quantum-chromodynamical crossover, at $T
\simeq 150$~MeV, drops to about 17.  The quantity $g_*^{1/3}$ thus drops from about 4.4
to 2.6, and we can see that the product $a T$ remains constant to a great extent. It is
then convenient to normalize the scale factor as the inverse of the temperature, $a = 1/
T$. With this choice, we have $a = \eta / M_*$, where $M_* = \left( 45 / 4 \pi^3 g_*
\right)^{1/2} M_{\rm P} \simeq 10^{18}$~GeV is the effective Planck mass, and $\sigma_c =
\sigma/ T \approx 70$ is constant in time \cite{Boyarsky:2011uy, Turner:1987bw,
Baym:1997gq}. The equilibrium spectrum (\ref{Seq}) is independent of time and, with this
normalization, reads
\beq \label{S0}
S_{\rm eq} (k, \eta) \equiv S_0 (k) = \frac{k}{e^k - 1} \, .
\eeq
The excess $\rho_B (\eta)$ of the energy density of magnetic field over the thermal
energy density is then determined by
\beq \label{rhoB}
\rho_B (\eta) = \frac{T^4}{2 \pi^2} \int_0^\infty P (k, \eta) k^2 d k \, ,
\eeq
where $P (k, \eta) \equiv S (k, \eta) - S_0 (k)$ is the excess over the thermal power
spectrum.

The system of equations (\ref{dotS}) and (\ref{dotA}) is supplemented by the evolution
equation for the chiral chemical potential \cite{Boyarsky:2011uy}:
\beq \label{muevo}
\frac{d \Delta \mu (\eta)}{d \eta} = - \frac{c_\Delta \alpha}{\pi^2} \int_0^\infty
\frac{\partial A (k, \eta)}{\partial \eta} k d k - \Gamma_{\rm f} (\eta) \Delta \mu
(\eta) + \beta (\eta) \, .
\eeq
Here, $c_\Delta$ is a numerical constant of order unity (it would be equal to $3/4$ in
pure quantum electrodynamics) that reflects the particle content of the primordial
plasma, $\beta (\eta)$ is an external source for the evolution of chiral chemical
potential which, for definiteness, we assume to be positive, and $\Gamma_{\rm f} (\eta)$
is the coefficient of the so-called `flipping' term which describes decay of lepton
chiral asymmetry caused by chirality flips in the electroweak interactions. 

\section{Development of helical magnetic fields}

Since the conformal conductivity $\sigma_c$ is constant during the period of evolution
under investigation, it is more convenient to work in terms of a rescaled conformal time
$\tau = \eta / \sigma_c$. With regard of (\ref{S0}), system (\ref{dotS}), (\ref{dotA})
can then be written in the form
\ber
\dot P  &=& - 2 k^2 P + \frac{2 \alpha k}{\pi} \Delta \mu A \, ,
\label{dotP1} \\
\dot A  &=& - 2 k^2 A + \frac{2 \alpha k}{\pi} \Delta \mu \left( P + S_0 \right) \, .
\label{dotA1}
\eer
Here and in what follows, the overdot denotes the derivative with respect to $\tau$. We
then set the initial conditions for system (\ref{muevo}), (\ref{dotP1}), (\ref{dotA1}) at
the formal cosmological singularity $\tau = 0$ in the form
\beq \label{inicon}
P (k, 0) \equiv 0 \, , \qquad A (k , 0) \equiv 0 \, , \qquad \gamma (0) = 0 \, .
\eeq
Hence, we also have $\Delta \mu (0) = 0$.  Thus, in the presence of lepton chiral
asymmetry ($\Delta \mu \ne 0$), generation of the helicity spectrum $A (k)$ commences,
according to (\ref{dotA1}), due to the presence of thermal distribution $S_0 (k)$.

Solution of system (\ref{dotP1}) and (\ref{dotA1}) with respect to the spectral functions
$P (k, \tau)$ and $A (k, \tau)$ with the initial conditions (\ref{inicon}) is given by
\ber
P (k, \tau) &=& \frac{2 \alpha k}{\pi} S_0 (k) e^{- 2 k^2 \tau} \int_0^\tau e^{2 k^2
\tau'} \sinh \left( 2 k \left[ \Psi (\tau) - \Psi (\tau') \right]
\right) \Delta \mu (\tau') d \tau' \, , \label{solP} \\
A (k, \tau) &=& \frac{2 \alpha k}{\pi} S_0 (k) e^{- 2 k^2 \tau} \int_0^\tau e^{2 k^2
\tau'} \cosh \left( 2 k \left[ \Psi (\tau) - \Psi (\tau') \right] \right) \Delta \mu
(\tau') d \tau' \, , \label{solA}
\eer
where
\beq \label{AB}
\Psi(\tau) = \frac{\alpha} {\pi} \int^{\tau}_{0} \Delta \mu (\tau') d \tau' \, .
\eeq
One can see that $A (k, \tau)$ has the same sign as $\Delta \mu (\tau)$, while $P (k,
\tau)$ is always positive.

\begin{figure}[htp]
\centerline{\includegraphics[width=.8\textwidth]{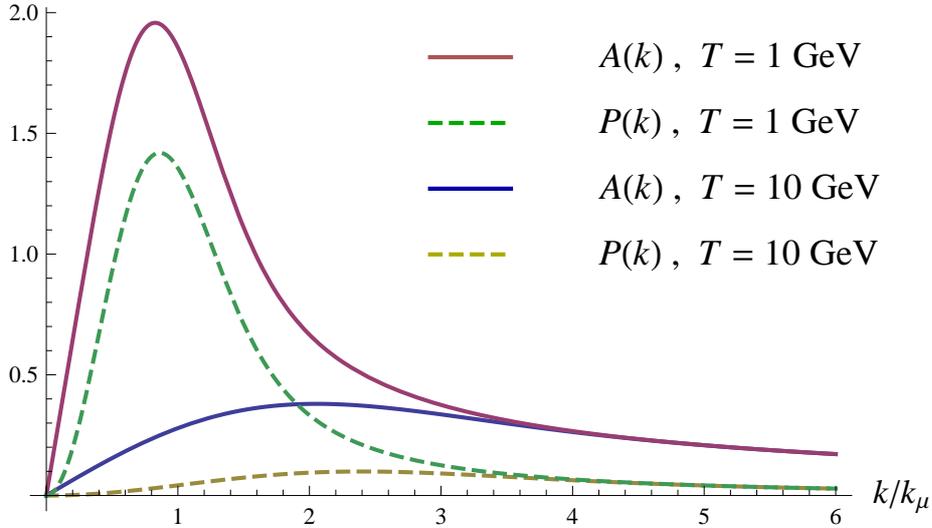}}
\caption{Spectrum (\ref{spec-P}) is plotted for $\Delta \mu = 5 \times 10^{-6}$ ($k_\mu
\simeq 10^{-8}$) and at the cosmological plasma temperatures $T = 10$~GeV (dashed yellow)
and $T = 1$~GeV (dashed green). Spectrum (\ref{spec-A}) is also plotted at $T = 10$~GeV
(solid blue) and $T = 1$~GeV (solid pink). The spectra increase in magnitude with time.
In the region $k / k_\mu \gg 1$, the spectra are saturated to their stationary
values $A (k) \simeq S_0 (k) k_\mu /k$ and $P (k) \simeq S_0 (k) \left(k_\mu
/k\right)^2$. \label{fig:spec}}
\end{figure}

To estimate the behavior of the spectral functions $P (k, \tau)$ and $A (k, \tau)$, let
us evaluate them under the condition $\Delta \mu \equiv {\rm const} > 0$ [which can be
ensured by an appropriate behavior of the source $\beta (\tau)$ in (\ref{muevo})]. Using
(\ref{AB}) and taking the elementary integrals in (\ref{solP}) and (\ref{solA}), we have
\ber \label{spec-P}
P (k, \tau) &=& \frac{S_0 (k)}{(k / k_\mu)^2 - 1}\left(1 - e^{- 2 k^2 \tau} \left[ \cosh
\left( 2 k_\mu k \tau \right) + \frac{k}{k_\mu} \sinh \left( 2 k_\mu k \tau \right)
\right] \right) \, , \\
A (k, \tau) &=& \frac{S_0 (k)}{(k / k_\mu)^2 - 1}\left(\frac{k}{k_\mu} - e^{- 2 k^2 \tau}
\left[ \sinh \left( 2 k_\mu k \tau \right) + \frac{k}{k_\mu} \cosh \left( 2 k_\mu k \tau
\right) \right] \right) \, , \label{spec-A}
\eer
where
\beq\label{kmu}
k_\mu = \frac{\alpha \Delta \mu}{\pi} \, .
\eeq
Spectra (\ref{spec-P}) and (\ref{spec-A}) for $\Delta \mu = 5 \times 10^{-6}$ ($k_\mu
\simeq 10^{-8}$) are plotted in figure~\ref{fig:spec} for temperatures $T = 10$~GeV and
1~GeV, corresponding to $\tau \approx 1.2 \times 10^{15}$ and $1.2 \times 10^{16}$,
respectively.

As can be seen from expressions (\ref{spec-P}) and (\ref{spec-A}) and from
figure~\ref{fig:spec}, there arise two characteristic regions of wavenumbers: the region
of relatively small $k$ (of order $k_\mu$), where the spectra keep growing and approach
the property $A (k) \simeq P (k) \gtrsim S_0 (k)$ of maximal helicity, and the region of
`tails' of these spectra, where they quickly reach the threshold values $A (k) \simeq S_0
(k) k_\mu /k$ and $P (k) \simeq S_0 (k) \left(k_\mu /k\right)^2$.

Indeed, in the region of large wavenumbers $k \gg k_\mu$, equation (\ref{dotA1}) is
approximated as
\beq
\dot A  \approx - 2 k^2 A + \frac{2 \alpha k}{\pi} \Delta \mu S_0 = - 2 k^2 A + 2 k_\mu k
S_0 \, ,
\eeq
with the solution
\beq
A (k) = S_0 (k) \frac{k_\mu}{k} \left( 1 - e^{- 2 k^2 \tau} \right) \, ,
\eeq
that exponentially with time approaches the equilibrium $A_{\rm eq} = S_0 k_\mu / k$.
Solution of (\ref{dotP1}) is then given by
\beq
P (k) = S_0 (k) \left( \frac{k_\mu}{k} \right)^2 \left[ 1 - e^{- 2 k^2 \tau} \left( 1 + 2
k^2 \tau \right) \right] \, ,
\eeq
with a rapid exponential convergence to the equilibrium $P_{\rm eq} = S_0 \left(k_\mu
/k\right)^2$.  For sufficiently slow evolution of $\Delta \mu (\tau)$, these expressions
for the spectral `tails' will retain their forms, with $k_\mu (\tau)$ expressed through
$\Delta \mu (\tau)$ by (\ref{kmu}).

It should be noted that the magnetohydrodynamical description of cosmic plasma cannot be
trusted in the domain of large physical wavenumbers $k / a \gtrsim \alpha T$, which, in
our system of conformal units, corresponds to values of $k \gtrsim \alpha \sim 10^{-2}$.
This limitation is insignificant for the evolution of magnetic instability developing on
much larger spatial scales (as is usually the case in the scenarios under consideration).
In the physically relevant domain $k \lesssim \alpha$, the thermal spectrum $S_0 (k)
\approx 1$. Therefore, we can replace the factor $S_0 (k)$ by unity in all our equations;
however, for the sake of better physical clarity, we will retain it.

In the region of relatively small wavenumbers, the spectra $P (k, \tau)$ and $A (k,
\tau)$ rapidly become equal to each other, as can be seen from figure~\ref{fig:spec}. In
the regime $P \gg S_0$, in which the quantity $S_0$ on the right-hand side of equation
(\ref{dotA1}) can be neglected, one can multiply equation (\ref{dotP1}) by $A (k, \tau)$,
equation (\ref{dotA1}) by $P (k, \tau)$ and subtract them to obtain an equation relating
the spectral functions:
\beq
\left( P^2 \right)^\cdot + 4 k^2 P^2 = \left( A^2 \right)^\cdot + 4 k^2 A^2 \, .
\eeq
This implies the relation
\beq\label{hellim}
P^2 (k , \tau) = A^2 (k, \tau) + f_0^2 (k) e^{- 4 k^2 \tau} \, ,
\eeq
where $f_0^2(k)$ is an integration constant.  We see that, if the quantity $A^2 (k,
\tau)$ is not decaying or decaying slower than the last exponent in (\ref{hellim}), then
a maximally helical state develops, with $P \simeq A$.  In this regime, system
(\ref{dotP1}), (\ref{dotA1}) reduces to a single equation for $A (k, \tau) \simeq P (k,
\tau)$:
\beq \label{helA}
\dot A  = \left( - 2 k^2 + \frac{2 \alpha k}{\pi} \Delta \mu \right) A \, .
\eeq

\section{Evolution of chiral asymmetry and magnetic field}

Assuming that a maximally helical configuration quickly develops at some initial time
$\tau_\ini$, we are going to establish how it will evolve together with $\Delta \mu
(\tau)$ after the source $\beta (\tau)$ in (\ref{muevo}) is switched off.

Let us make the notation $\Delta \mu_\ini = \Delta \mu (\tau_\ini)$, $P_\ini (k) = P (k,
\tau_\ini)$, $A_\ini (k) = A (k, \tau_\ini)$, and introduce the momentum $k_\ini$
similarly to (\ref{kmu}):
\beq\label{kini}
k_\ini = \frac{\alpha \Delta \mu_\ini}{\pi} \, .
\eeq
The initial spectra can be presented as
\beq \label{APini}
P_\ini (k) = P_0 Z \left(\frac{k}{k_\ini} \right) \, , \qquad A_\ini (k) \approx P_\ini
(k) \left( 1 + \frac{k}{k_\ini} \right) \, ,
\eeq
where $Z (x) $ describes the shape of the spectrum, and the factor $(1 + k / k_\ini)$ is
introduced to reflect the relation in the `tails' of the spectra. The normalization
constant $P_0$ is chosen so that
\beq \label{normal}
\int_0^\infty Z (x) x^2 d x = 1 \, ,
\eeq
and the initial excess (\ref{rhoB}) of the energy density of magnetic field over the
thermal energy density is then equal to
\beq \label{helrhoB}
\rho_B (\tau_\ini) = \frac{T_\ini^4}{2 \pi^2} \int_0^\infty P (k, \tau_\ini) k^2 d k =
\frac{T_\ini^4 k_\ini^3 P_0}{2 \pi^2} \, .
\eeq
It is convenient to relate this quantity to the total radiation energy density by
introducing the dimensionless parameter
\beq \label{rB}
r_B^\ini \equiv \frac{\rho_B (\tau_\ini)}{\rho_r (\tau_\ini)} = \frac{30 \rho_B
(\tau_\ini)}{\pi^2 g_* T_\ini^4}  = \frac{15}{2 \pi^4 g_*} P_0 k_\ini^3 \, .
\eeq

Asymptotically, as $\tau \to \infty$, the maximum of spectrum (\ref{spec-P}) is reached
at $k = k_\mu / 2$, and one can derive an approximate asymptotic estimate for $r_B^\ini$:
\beq
r_B^\ini \simeq \frac{k_\ini^2}{\pi^4} \sqrt{\frac{1}{2 \tau_\ini}}\, e^{k_\ini^2
\tau_\ini / 2} \, ,
\eeq
where $k_\ini$ is given by (\ref{kini}).  One can see the exponential dependence of
$r_B^\ini$ on the (rescaled) conformal time $\tau_\ini$ (or temperature $T_\ini = M_* /
\sigma_c \tau_\ini$) at which the spectrum (\ref{spec-P}) is finally developed by the
external source of chiral asymmetry. In what follows, we take $r_B^\ini$ and $\Delta
\mu_\ini$ [related to $k_\ini$ by (\ref{kini})] to be our independent parameters.

The subsequent evolution of the spectrum in the domain where $P (k, \tau) \approx A (k,
\tau) \gg S_0 (k)$ is described by equation (\ref{helA}). Its solution with the initial
condition $A (k, \tau_\ini) = A_\ini (k)$ is given by
\beq
A (k, \tau) = g^2 (k, \tau) A_\ini (k) \, ,
\eeq
where
\beq \label{g}
g (k, \tau) = e^{- k^2 \Delta \tau  + k \Delta \Psi (\tau)} \, , \quad \Delta \tau = \tau
- \tau_\ini \, , \quad \Delta \Psi (\tau) = \Psi (\tau) - \Psi (\tau_\ini) \, ,
\eeq
is the growth factor.

Solution of (\ref{muevo}) with the zero source $\beta$ and with the flipping term
neglected can be written as
\beq \label{differ}
\Delta \mu_\ini - \Delta \mu (\tau) = \frac{c_\Delta \alpha}{\pi^2} \int_0^\infty \left[
A (k, \tau) - A_\ini (k) \right] k d k \, .
\eeq
The contribution of the `tail' in the distributions $A (k, \tau)$ and $A_\ini (k)$ to the
value of the difference $\Delta \mu_\ini - \Delta \mu (\tau)$ in this expression is
negligibly small. Indeed, the integral over the `tail' region is estimated as
\ber
\frac{c_\Delta \alpha}{\pi^2} \int_{k_{\rm tail}}^\infty \left[ A (k, \tau) - A_\ini (k)
\right] k d k &\simeq& \frac{c_\Delta \alpha^2}{\pi^3} \left[ \Delta \mu (\tau) - \Delta
\mu_\ini \right] \int_{k_{\rm tail}}^\infty S_0 (k) d k \nonumber \\
&\simeq& \frac{c_\Delta \alpha^2}{\pi^3} \left[ \Delta \mu (\tau) - \Delta \mu_\ini
\right] \, ,
\eer
which is much smaller by absolute value than the left-hand side of (\ref{differ}) because
of the smallness of $\alpha \approx 1/137$. Thus, we can ignore the presence of power-law
tails in the spectra in (\ref{differ}), and write, using (\ref{APini}),
\beq
\Delta \mu_\ini - \Delta \mu (\tau) = \frac{c_\Delta \alpha k_\ini^2 P_0}{\pi^2}
\int_0^\infty \left[ g^2 (k_\ini x, \tau) - 1 \right] Z (x) x d x \, .
\eeq
Dividing this by $\Delta \mu_\ini$ and using (\ref{kini}) and (\ref{rB}), we obtain the
estimate
\beq \label{solmu}
1 - \frac{\Delta \mu (\tau)}{\Delta \mu_\ini} = \frac{\pi c_\Delta \alpha^2 g_*
r_B^\ini}{15 k_\ini^2} \int_0^\infty \left[ g^2 (k_\ini x, \tau) - 1 \right] Z (x) x d x
\, .
\eeq

For values in a broad typical range of parameters in different cosmological scenarios,
the factor in front of the integral in (\ref{solmu}) is much larger than unity.  For
instance, for $g_* = 75$ and $\Delta \mu_\ini = 5 \times 10^{-6}$, this factor is
estimated to be $\sim 10^{13} r_B^\ini$, and will be very large for $r_B^\ini \gg
10^{-13}$. Since the left-hand side of (\ref{solmu}) is always bounded by unity, this
implies that the integral on the right-hand side should be extremely small.  The relation
\beq \label{fincon}
\int_0^\infty \left[ g^2 (k_\ini x, \tau) - 1 \right] Z (x) x d x \approx 0
\eeq
can then be regarded as an integral equation implicitly expressing the quantity to be
found $\Delta \Psi$ through the known quantity $\Delta \Phi$ [both enter the function
$g^2 (k_\ini x, \tau)$ under this integral; see (\ref{g})].

It is convenient to introduce the variables
\beq \label{phi}
\phi = 2 k_\ini^2 \Delta \tau \, , \qquad \psi = 2 k_\ini \Delta \Psi \, .
\eeq
In terms of these variables, we have
\beq \label{g2}
g^2 (k_\ini x, \tau) = e^{- \phi x^2 + \psi x} \, ,
\eeq
and equation (\ref{fincon}) establishes the dependence $\psi (\phi)$, which is determined
only by the form  $Z (x)$ of the initial distribution.  To find the dependence $ \psi
(\phi)$, we differentiate (\ref{fincon}) with respect to $\phi$. We obtain the Cauchy
problem
\beq \label{dBA}
\frac{d \psi}{d \phi} = \frac{\int_0^\infty e^{- \phi x^2 + \psi x} Z(x) x^3 d
x}{\int_0^\infty e^{- \phi x^2 + \psi x} Z(x) x^2 d x} \, , \qquad \psi (0) = 0 \, .
\eeq
The evolution of the chiral chemical potential can then be calculated by using
(\ref{AB}):
\beq \label{muev}
\frac{\Delta \mu }{\Delta \mu_\ini} = \frac{\pi}{\alpha \Delta \mu_\ini} \dot \Psi =
\psi'(\phi) \, .
\eeq
Remarkably, the evolution of the chiral chemical potential $\Delta \mu$ depends on the
initial conditions and on time through a single scaling parameter $\phi$, defined in
(\ref{phi}).

\begin{figure}[htp]
\centerline{\includegraphics[width=.8\textwidth]{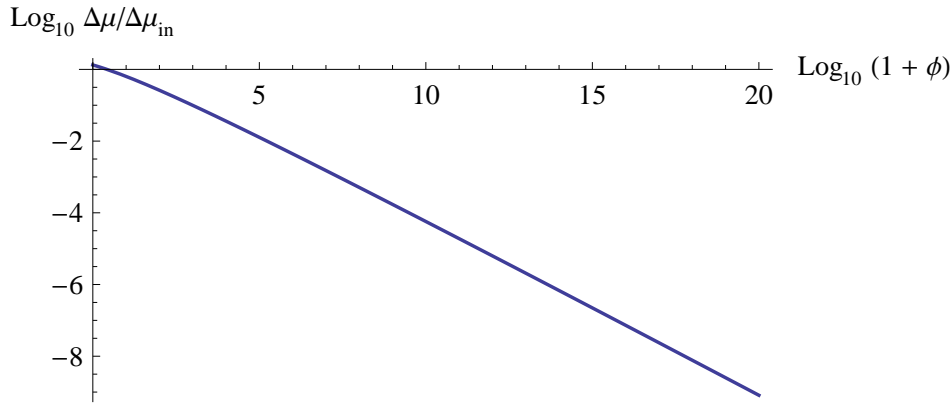}}
\caption{Evolution (\ref{muevol}) of the chiral chemical potential plotted in logarithmic scale.
\label{fig:muevol}}
\end{figure}

As an example, let us approximate the initial spectrum $P_\ini (k)$ with subtracted
high-frequency `tail' by a corresponding normalized spectral function with exponential
cut-off:\footnote{Equation (\ref{zx}) gives a correct linear growth at small $x$,
observed in figure~\ref{fig:spec}.}
\beq \label{zx}
Z (x) = 2 x e^{- x^2} \, .
\eeq
Then, introducing the variable
\beq \label{zeta}
\zeta = \frac{\psi}{\sqrt{1 + \phi}} \, ,
\eeq
we can present problem (\ref{dBA}) in the form
\beq \label{dzeta}
\frac{d \zeta}{d \phi} = \frac{F (\zeta) - \frac12 \zeta}{1 + \phi} \, , \qquad \zeta (0)
= 0 \, ,
\eeq
where
\beq \label{F}
F (\zeta) = \frac{\int_0^\infty e^{- x^2 + \zeta x} x^4 d x}{\int_0^\infty e^{- x^2 +
\zeta x} x^3 d x} \, .
\eeq
Equation (\ref{dzeta}) can, in principle, be integrated, and the function $\zeta (\phi)$
can be found. The evolution (\ref{muev}) of the chiral chemical potential is then given
by
\beq \label{muevol}
\frac{\Delta \mu }{\Delta \mu_\ini} = \psi'(\phi) = \frac{F \left(\zeta (\phi)
\right)}{\sqrt{1 + \phi}} \, ,
\eeq
where $F (\zeta)$ is given by (\ref{F}).  Solution (\ref{muevol}) is plotted in
logarithmic scale in figure~\ref{fig:muevol}.

Let us establish the late-time asymptotics of the solution to (\ref{dzeta}), (\ref{F}).
In the regime $\zeta \gg 1$, we have $F (\zeta) \approx \zeta/2 + 3/\zeta + {\cal O}
\left( \zeta^{-3} \right)$. Solution of (\ref{dzeta}) in this case behaves as
\beq \label{large-a}
\zeta (\phi) \approx \left[ {\rm const} + 6 \log (1 + \phi) \right]^{1/2} \, , \qquad
\phi \gg 1 \, .
\eeq
This qualitative behavior does not depend on the specific shape (\ref{zx}) of the initial
spectrum and is caused by the inverse cascade that transfers the spectral power to
small-frequency region.  Indeed, for large enough values of $\tau$, function (\ref{g2})
develops a strong Gaussian peak at small values of $x$, where $Z (x)$ behaves rather
smoothly (typically, as a power of $x$). Expression (\ref{dBA}) does not then depend on
the concrete form of $Z (x)$ in this limit.

\begin{figure}[htp]
\centerline{\includegraphics[width=.6\textwidth]{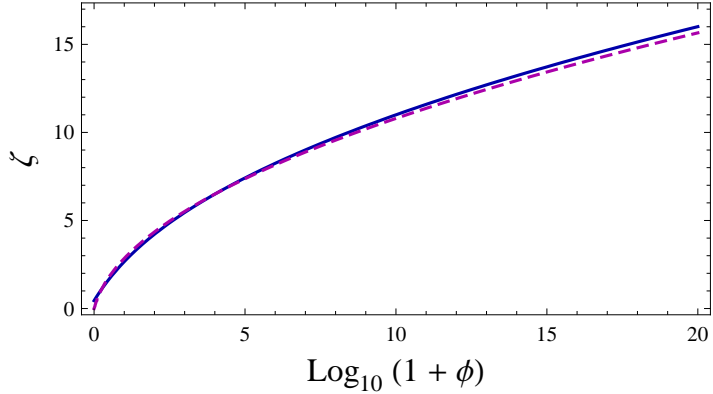}}
\caption{Exact solution of system (\ref{dzeta}), (\ref{F}) (solid blue line) versus interpolation
(\ref{zeta1}) (dashed pink line). \label{fig:interpol}}
\end{figure}

With the account of asymptotics (\ref{large-a}), the solution $\zeta (\phi)$ of the
differential equation (\ref{dzeta}), (\ref{F}) can be approximated by the expression
\beq \label{zeta1}
\zeta (\phi) = \left[ \sqrt{1 + 6 \log (1 + \phi)} - 1 \right] \, .
\eeq
Numerical integration confirms this approximation within about $1\%$ precision (see
figure~\ref{fig:interpol}). Then
\beq \label{interpol}
\psi (\phi) = \sqrt{1 + \phi}\, \zeta (\phi) = \sqrt{1 + \phi} \left[ \sqrt{1 + 6 \log (1
+ \phi)} - 1 \right] \, .
\eeq

Using (\ref{muevol}), we then obtain the universal late-time asymptotics
\beq \label{muas}
\log \frac{\Delta \mu }{\Delta \mu_\ini} \approx - \frac12 \log (1 + \phi) + \frac12 \log
\log (1 + \phi) \, .
\eeq
This describes very well the almost ideal power-law behavior observed in
figure~\ref{fig:muevol} for large values of $\phi$, with the asymptotic power index equal
to $- 1/2$. The late-time asymptotics $\Delta \mu \propto \eta^{-1/2}$ in the system
under consideration has been previously established in \cite{Hirono:2015rla} (and in
\cite{Xia:2016any} with a leading logarithmic correction, $\Delta \mu \propto \eta^{-1/2}
\log^{1/2} \eta$), and also noted in \cite{Yamamoto:2016xtu} in the context of chiral
magnetohydrodynamics.

Let us also determine the behavior of the magnetic-field energy density, described by the
parameter $r_B (\tau)$.  For our developed chiral distribution, we have $P (k, \tau)
\approx g^2 (k, \tau) P_\ini (k)$. Hence,
\beq \label{NB}
r_B (\tau) = N_B^{}  r_B^\ini \, , \qquad N_B (\phi)  = \int_0^\infty Z (x) e^{- \phi x^2
+ \psi (\phi) x} x^2 d x  \, .
\eeq
Thus, the energy density of magnetic field also depends on time through a single scaling
parameter $\phi$, defined in (\ref{phi}).  Its behavior at large values of $\phi$ will
depend only on the behavior of the function $Z (x)$ at small $x$.

\begin{figure}[ht]
\begin{minipage}[t]{.48\textwidth}
    \vspace{0pt} \includegraphics[width=\textwidth]{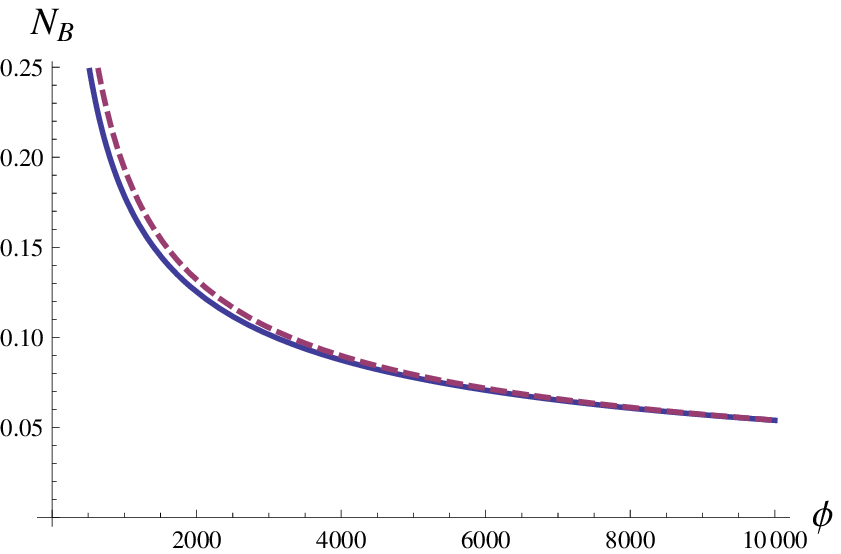}
  \end{minipage}
\begin{minipage}[t]{.51\textwidth}
    \vspace{0pt} \includegraphics[width=\textwidth]{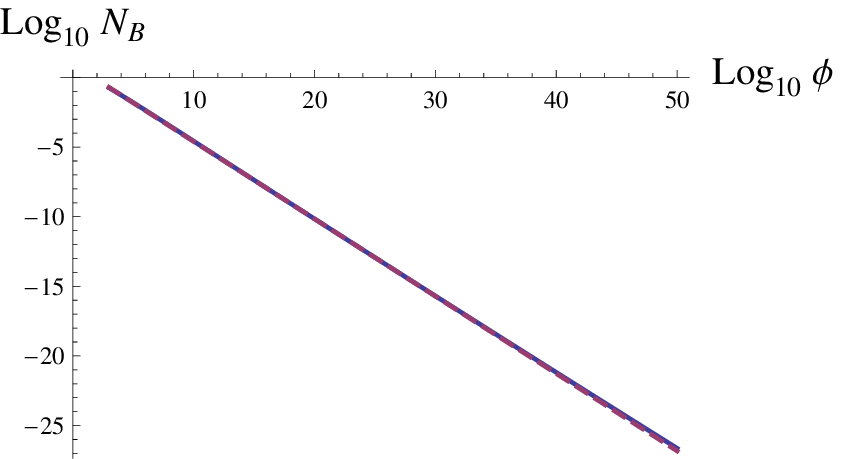}
\end{minipage}
\caption{The function $N_B (\phi)$ in (\ref{NB}) with $Z (x)$ given by (\ref{zx}) (solid
blue line) and its interpolation by (\ref{interpol-1}) (dashed pink line) in different
ranges of argument.  The curves on the right graph practically coincide.
\label{fig:interpol-1}}
\end{figure}

For the initial shape (\ref{zx}) of the magnetic-field spectrum, the scaling function
$N_B (\phi)$ is determined by approximations (\ref{zeta1}) and (\ref{interpol}). Analytic
estimate of integral (\ref{NB}) gives a rather complicated asymptotics at $\phi \gg 1$\,:
\beq
N_B (\phi) \propto \frac{\log^{3/2} \phi}{\sqrt{\phi}} \, e^{- \frac12 \sqrt{1 + 6 \log
\phi}} \, .
\eeq
However, in a very wide range of the values of argument, $10^2 \lesssim \phi \lesssim
10^{50}$, function (\ref{NB}) is excellently interpolated by a simple power law (see
figure~\ref{fig:interpol-1})
\beq \label{interpol-1}
N_B (\phi) \approx 9 \phi^{- 5/9} \, .
\eeq

With $\phi$ being asymptotically given by [see (\ref{phi}) and (\ref{kini})]
\beq \label{phi-as}
\phi = \frac{2 \alpha^2 M_* \Delta \mu_\ini^2}{\pi^2 \sigma_c T} \, ,
\eeq
equations (\ref{muas}) and (\ref{NB}), (\ref{interpol-1}) give the asymptotic behavior of
the quantities $\Delta \mu$ and $r_B$ as functions of temperature and of their initial
values:
\beq
\frac{\Delta \mu }{\Delta \mu_\ini} = \left[ \frac{\pi^2 \sigma_c T}{2 \alpha^2 M_*
\Delta \mu_\ini^2} \log \frac{2 \alpha^2 M_* \Delta \mu_\ini^2}{\pi^2 \sigma_c T}
\right]^{1/2} \, , \qquad \frac{r_B}{r_B^\ini} = 9 \left( \frac{\pi^2 \sigma_c T}{2
\alpha^2 M_* \Delta \mu_\ini^2} \right)^{5/9}
\eeq

\section{Decay of chiral asymmetry caused by chirality flipping}

In the preceding analysis, we totally neglected the flipping term with coefficient
$\Gamma_{\rm f} (\eta)$ in (\ref{muevo}), which is justified at high temperatures.
However, as the temperature drops down because of cosmological expansion, at some point
this term starts dominating over the other terms on the right-hand side of (\ref{muevo}),
after which the chiral chemical potential decays exponentially as
\beq \label{expodec}
\Delta \mu (T) \propto \exp \left( - \frac{\alpha^2 M_* m_e^2}{27\, T^3} \right) \, .
\eeq
In this section, we estimate the temperature $T_{\rm f}$ at which this decay commences.

The contribution to the coefficient $\Gamma_{\rm f} (\tau)$ comes from weak and
electromagnetic processes, so that we have $\Gamma_{\rm f} = \Gamma_w + \Gamma_e$, where
the weak and electromagnetic contributions are estimated, respectively, as (see
\cite{Boyarsky:2011uy})
\beq \label{gammas}
\Gamma_w \sim G_{\rm F}^2 T^4 \left( \frac{m_e}{3 T} \right)^2 \, , \qquad \Gamma_e \sim
\alpha^2 \left( \frac{m_e}{3 T} \right)^2 \, .
\eeq
Here, $G_{\rm F}$ is the Fermi constant, $\alpha$ is the fine structure constant, and
$m_e$ is the electron mass.  The factors in the brackets in (\ref{gammas}) with electron
mass $m_e$ describe suppression of chirality-flipping scattering rates with respect to
`chirality-preserving' ones.  The weak contribution $\Gamma_w$ dominates at temperatures
$T > T_{\rm eq} \simeq \sqrt{\alpha/ G_{\rm F}} \approx 25$~GeV, while, at $T < T_{\rm
eq}$, chirality flipping is dominated by the electromagnetic processes and is
characterized by $\Gamma_e$ in (\ref{gammas}).

In view of equation (\ref{dotA1}), the first term on the right-hand side of (\ref{muevo})
is itself a sum of two terms with opposite signs:
\beq \label{two-term}
- \frac{c_\Delta \alpha}{\pi^2 \sigma_c} \int_0^\infty \dot A (k, \tau) k d k = \frac{2
c_\Delta \alpha}{\pi^2 \sigma_c} \int_0^\infty A (k, \tau) k^3 d k - \frac{2 c_\Delta
\alpha^2 \Delta \mu}{\pi^3 \sigma_c} \int_0^\infty \left[ P (k, \tau) + S_0 (k) \right]
k^2 d k \, .
\eeq
In this expression, the thermal `tail' in the spectral function $A (k, \tau)$ compensates
the thermal contribution from $S_0 (k)$.  Indeed, at the `tail,' we have $A = A_{\rm eq}
= S_0 \left(k_\mu/ k \right)$, and taking into account (\ref{kmu}), we observe
cancellation of the corresponding integrals in (\ref{two-term}).  Therefore, it is the
negative term with the integral of the spectral function $P (k, \tau)$ in
(\ref{two-term}) that is to be considered. To determine whether the neglect of the
flipping term in (\ref{muevo}) is legitimate, we should compare the absolute value of
this term with the absolute value $\Gamma_{\rm f} (\tau) \Delta \mu (\tau)$ of the
flipping term. It is convenient to divide both quantities by $\Delta \mu (\tau)$. For the
first expression, we have
\beq \label{regdec}
\frac{2 c_\Delta \alpha^2}{\pi^3 \sigma_c} \int_0^\infty P (k, \tau) k^2 d k = \frac{2
c_\Delta \alpha^2}{\pi^3 \sigma_c} g_* r_B (\tau) \int_0^\infty S_0 (k) k^2 d k = \frac{2
\pi c_\Delta \alpha^2}{15\, \sigma_c} g_* r_B (\tau)\, ,
\eeq
Using (\ref{NB}) and (\ref{interpol-1}) to express $r_B (\tau)$ through $r_B^\ini$, we
estimate (\ref{regdec}) as
\beq \label{regdec1}
\frac{2 \pi c_\Delta \alpha^2}{15\, \sigma_c} g_* r_B (\tau) \simeq \frac{\pi \alpha^2
}{\sigma_c \phi^{5/9}}  g_* r_B^\ini \, .
\eeq
This expression should be compared to each of the quantities in (\ref{gammas}).  Assuming
that chirality flipping comes into play at temperatures $T < T_{\rm eq} \approx 25$~GeV
(this will be confirmed by the final estimate), we only need to take into account the
electromagnetic part $\Gamma_e$. We then have an equation for the estimate of the
temperature of decay caused by chirality flipping:
\beq \label{finest}
\frac{\pi}{\sigma_c \phi^{5/9}}  g_* r_B^\ini \simeq \left( \frac{m_e}{3 T} \right)^2 \,
.
\eeq
The asymptotic value of $\phi \gg 1$ is given by (\ref{phi-as}).  Substituting it into
(\ref{finest}) and solving the resulting equation with respect to $T$, we obtain
\beq \label{Tf}
T_{\rm f} =  \left( \frac{2^5 \alpha^{10} \sigma_c^4 m_e^{18} M_*^5 \Delta
\mu_\ini^{10}}{3^{18} \pi^{19} \left[ g_* r_B^\ini \right]^9} \right)^{1/23} \approx
\frac{1.6 \times 10^3}{\sqrt{g_*}} \, \left( \frac{\Delta \mu_\ini^{10}}{\left[ r_B^\ini
\right]^9} \right)^{1/23} \, \, \mbox{MeV} \, ,
\eeq
where we have put the numerical values for physical constants. For $g_* = 30$, $\Delta
\mu_\ini = 3 \times 10^{-5}$, and $r_B^\ini = 5 \times 10^{-5}$, this equation gives
$T_{\rm f} \simeq 150$~MeV (at this time, $\phi \approx 1200$). For $r_B^\ini = 5 \times
10^{-4}$, we obtain $T_{\rm f} \simeq 60$~MeV (with $\phi \approx 3000$). This is in good
qualitative agreement with the numerical results of \cite{Boyarsky:2011uy}.

Note that the resulting temperature (\ref{Tf}) does not depend on the temperature at
which the initial values $\Delta \mu_\ini$ and $r_B^\ini$ are set (and which is assumed
to be much higher than $T_{\rm f}$).  This is due to the asymptotic scaling $\Delta \mu
\propto \phi^{-1/2}$ [see (\ref{muas})], and $r_B \propto \phi^{-5/9}$ [see (\ref{NB})
and (\ref{interpol-1})], ensuring that the ratio $\Delta \mu^{10}/ r_B^9$ remains to be
roughly constant in the regime of inverse cascade.

\section{Summary}

We provided an analytic treatment of the process of generation of helical magnetic field
in an early hot universe in the presence of externally induced lepton chiral asymmetry,
and of the subsequent mutual evolution of the chiral asymmetry and magnetic field.
Helical magnetic field is generated from the thermal initial spectrum (extrapolated to
all scales including the super-horizon ones) owing to the effects of quantum chiral
anomaly. The thermal bath also serves as a medium of relaxation of magnetic field to its
thermal state.  The generated helical magnetic field and the lepton chiral asymmetry are
capable of supporting each other, thus prolonging their existence to cosmological
temperatures as low as tens of MeV, with spectral power being permanently transformed
from small to large spatial scales (the so-called `inverse cascade')
\cite{Boyarsky:2011uy}.

Our main results are summarized as follows. We obtained analytic expressions describing
the evolution of the lepton chiral chemical potential and magnetic-field energy density.
For a developed maximally helical magnetic field, both the chiral chemical potential
$\Delta \mu$ and the relative fraction of magnetic-field energy density $r_B$ depend on
their initial values and on time through a single variable $\phi$ introduced in
(\ref{phi}). This scaling property is encoded in equations (\ref{dzeta})--(\ref{muevol})
and (\ref{NB})--(\ref{interpol-1}), and depicted in figures \ref{fig:muevol} and
\ref{fig:interpol-1}. The late-time asymptotics for $\Delta \mu$ reproduces the scaling
law $\Delta \mu \propto \eta^{-1/2} \log^{1/2} \eta$ [see equation (\ref{muas})]
previously found in this system in \cite{Hirono:2015rla, Xia:2016any}. By numerical
interpolation, we find that the relative fraction $r_B$ of the magnetic-field energy
density in this regime decays as $r_B \propto \eta^{-5/9}$ all through the relevant part
of the cosmological history. Since the conformal time $\eta$ in our units is related to
the temperature $T$ as $\eta = M_* / T$, this also describes the evolution of these
quantities with temperature.

As the temperature drops to sufficiently low values due to the cosmological expansion,
the chirality-flipping lepton scattering processes take control over the evolution of
chiral asymmetry, leading to its rapid decay (\ref{expodec}).  We derived a simple
expression (\ref{Tf}) for the temperature at which this happens, depending on the
initially generated values of the energy density of magnetic field and of the lepton
chiral asymmetry.

The analytic expressions obtained in this paper are sufficiently general and may be used
for primary evaluation of scenarios of cosmological magnetogenesis by lepton chiral
asymmetry.

\acknowledgments

We are grateful to Alexey Boyarsky and Oleg Ruchayskiy for valuable comments. M.~S. and
O.~T. acknowledge support from the Scientific and Educational Center of the Bogolyubov
Institute for Theoretical Physics. The work of O.~T. was supported by the WFS National
Scholarship Programme and by the Deutsche Forschungsgemeinschaft (DFG) in part through
the Collaborative Research Center ``The Low-Energy Frontier of the Standard Model'' (SFB
1044), in part through the Graduate School ``Symmetry Breaking in Fundamental
Interactions'' (DFG/GRK 1581), and in part through the Cluster of Excellence ``Precision
Physics, Fundamental Interactions and Structure of Matter'' (PRISMA). The work of Y.~S.
was supported by the Swiss National Science Foundation grant SCOPE IZ7370-152581.

\end{document}